\documentclass[a4paper, 11pt]{article}

\pdfoutput=1 

\usepackage{amsmath}
\usepackage[T1]{fontenc} 
\usepackage{authblk}
\usepackage[a4paper, total={6in, 8in}]{geometry}
\usepackage{graphicx}

\begin{document} 

\title{\bf Weak charges in \boldmath $SU(5)_{L} \times U(1)_{Y}$ \bf{gauge models}}

\author{Adrian Palcu}
\affil{"Aurel Vlaicu" University of Arad,\\ 2 Elena Dr\u{a}goi Street, Arad-310330, Romania}
\date{}

\maketitle

\abstract{Within the framework of a renormalizable $SU(5)_{L} \times U(1)_{Y}$ electro-weak gauge model with no exotic electric charges, we obtain all the neutral weak charge operators and their quantization, once the diagonalization of the neutral boson mass matrix is properly performed. Our results open up the path to a rich and promising phenomenological outcome. All the Standard Model phenomenology is recovered by simply decoupling the latter's scale ($v_{SM}=246$ GeV) from the higher scale ($V\sim$ 10 TeV) specific to our new electro-weak unification.}

\section{Introduction}
\label{sec:1}
In a recent paper \cite{1} the author proposed an original $SU(5)_{L} \times U(1)_{Y}$ gauge symmetry for the electro-weak unification, assuming the fact that the Standard Model (SM) \cite{2}-\cite{4} - based on the $SU(2)_{L} \times U(1)_{Y}$ electro-weak group - must obviously be somehow extended in order to properly address some experimental challenges such as the dark matter puzzle, the neutrino oscillation phenomenon (implying tiny, but massive, neutrinos), the particular electric charge quantization observed in nature, the number of precisely 3 fermion generations, the recently reported muon g-2 discrepancies, etc. The present letter simply aims at obtaining the neutral charges - computed as the couplings of the fermion fields to specific vector gauge bosons that mediate the weak interactions in the above mentioned $SU(5)_{L} \times U(1)_{Y}$ model. Consequently, we argue that the outcome is viable from phenomenological standpoint. Our approach relies on the method for treating generalized $SU(n)_{L} \times U(1)_{Y}$ gauge models with spontaneous symmetry breaking (SSB), conceived some years ago by Cot\u{a}escu \cite{5} and recently developed further by the author \cite{6}.   

The paper is organized in five sections, each dealing with some particular aspects of the extended SM to the $SU(3)_{c} \times SU(5)_{L} \times U(1)_{Y}$  gauge group (in short 3-5-1 model). In Section2 we briefly review the particle content of the model under consideration here (leptons and quarks irreducible representations), the gauge fields (with a focus on the interactions mediated by the four neutral vector bosons $Z$, $Z^{\prime}$, $Z^{\prime\prime}$, $Z^{\prime\prime\prime}$) and the scalar sector (responsible - via a specific Higgs mechanism - for the SSB). Section 3 deals properly with the mass matrix of the above mentioned four neutral vector bosons and its diagonalization that finally supplies - by employing the appropriate $\omega \in SO(4)$ matrix - the corresponding neutral charge operators. In Section 4 these operators are computed in detail for all fermion representations and, hence, their precise quantization is obtained. Section 5 is reserved for some concluding remarks.         
  
\section{\boldmath $SU(5)_{L} \times U(1)_{Y}$ gauge model}
\label{sec:2}

The particle content of the 3-5-1 model at hand is displayed below. There are three left-handed generations of leptons and quarks (see Ref.\cite{1}), occurring in the following distinct left-handed quintuplets:

\begin{equation}
L_{iL}=\left(\begin{array}{c}
N_{i}^{\prime\prime}\\
N_{i}^{\prime}\\
N_{i}\\
\nu_{i}\\
e_{i}
\end{array}\right)_{L}, \quad Q_{1L}=\left(\begin{array}{c}
U^{\prime\prime}\\
U^{\prime}\\
U\\
u\\
d
\end{array}\right)_{L}  , \quad Q_{2L,3L}=\left(\begin{array}{c}
D_{2,3}^{\prime\prime}\\
D_{2,3}^{\prime}\\
D_{2,3}\\
d_{2,3}\\
u_{2,3}
\end{array}\right)_{L},
\end{equation}
with $i=1,2,3$, and the corresponding right-handed singlet partners.
 
Their irreducible representations with respect to the model's whole gauge group $SU(3)_{c} \times SU(5)_{L} \times U(1)_{Y}$ are summarized below:
\begin{equation}
L_{iL}\sim(\boldsymbol{1},\boldsymbol{5},-\frac{1}{5}) \quad , \quad e_{iR}\sim(\boldsymbol{1},\boldsymbol{1},-1) \quad ,\quad \nu_{iR},N_{iR},N_{iR}^{\prime},N_{iR}^{\prime\prime}\sim(\boldsymbol{1},\boldsymbol{1},0)
\end{equation}

\begin{equation}
Q_{1L}\sim(\boldsymbol{3},\boldsymbol{5},\frac{7}{15}) \quad , \quad Q_{kL}\sim(\boldsymbol{3},\boldsymbol{5^{*}},-\frac{2}{15})
\end{equation}

\begin{equation}
u_{kR},u_{R},U_{R},U_{R}^{\prime},U_{R}^{\prime\prime}\sim(\boldsymbol{3},\boldsymbol{1},\frac{2}{3})\quad , \quad d_{R},d_{kR}, D_{kR},D_{kR}^{\prime},D_{kR}^{\prime\prime}\sim(\boldsymbol{3},\boldsymbol{1},-\frac{1}{3})
\end{equation}
with $k=2,3$.

These assignments are not arbitrary at all, but inferred \cite{1,6} by imposing the renormalization criteria that require for all the axial anomalies to be canceled. The general method \cite{5} was proved to predict - when this strict requirement is fulfilled - (i) precisely 3 fermion generations \cite{6} (if the number of colors in the $SU(3)_{c}$ of the QCD is kept, as usual, to 3) and (ii) the electric charge quantization \cite{6} observed in nature, as well. This seems to be a common result of many sorts of SM extensions emerging at not very high scales above TeV threshold. For example, $SU(3)_{L}\times U(1)_{Y}$ gauge models with \cite{7}-\cite{12} or without \cite{13}-\cite{26} exotic electric charges or (more recently) those based on the $SU(4)_{L}\times U(1)_{Y}$ gauge group \cite{27}-\cite{51} have been considered as plausible scenarios for particle physics. The richer phenomenology all these models exhibit and can predict has been extensively discussed in the literature. Yet - to our best knowledge - there is still no attempt to properly address the larger possible \cite{6} SM-extension, namely the one based on the electro-weak $SU(5)_{L}\times U(1)_{Y}$ gauge group, except for a short letter \cite{52} considering only a very particular such model involving exotic electric charges. As we have already stated it, our approach goes differently and considers only the case with no exotic electric charges, in realistic connection to the experimental observations to date.     
 
The electro-weak interactions in the model are mediated by the vector bosons supplied by the adjoint representation of the semi-simple gauge group employed above, namely 
\begin{equation}
A_{\mu}=\left(\begin{array}{ccccccccc}
D_{\mu}^{1} &  & Y_{\mu}^{\prime\prime0} &  & Y_{\mu}^{\prime0} &  & Y_{\mu}^{0} &  & Y_{\mu}^{\prime\prime+}\\
\\
Y_{\mu}^{\prime\prime0*} &  & D_{\mu}^{2} &  & X_{\mu}^{\prime\prime0} &  & X_{\mu}^{\prime0} &  & Y_{\mu}^{\prime+}\\
\\
Y_{\mu}^{\prime0*} &  & X_{\mu}^{\prime\prime0*} &  & D_{\mu}^{3} &  & X_{\mu}^{0} &  & Y_{\mu}^{+}\\
\\
Y_{\mu}^{0*} &  & X_{\mu}^{\prime0*} &  & X_{\mu}^{0*} &  & D_{\mu}^{4} &  & W_{\mu}^{+}\\
\\
Y_{\mu}^{\prime\prime-} &  & Y_{\mu}^{\prime-} &  & Y_{\mu}^{-} &  & W_{\mu}^{-} &  & D_{\mu}^{5}
\end{array}\right),
\end{equation}
with the diagonal entries (corresponding to the Cartan subalgebra of the $su(5)_{L}\times u(1)_{Y}$ algebra) considered in order as:
 
\begin{equation}
\begin{array}{ll}
D_{\mu}^{1} =&  \frac{1}{2}A_{\mu}^{3}+\frac{1}{2\sqrt{3}}A_{\mu}^{8}+\frac{1}{2\sqrt{6}}A_{\mu}^{15}+\frac{1}{2\sqrt{10}}A_{\mu}^{24}+YB^{0}_{\mu} \\
\\
D_{\mu}^{2} =  &  -\frac{1}{2}A_{\mu}^{3}+\frac{1}{2\sqrt{3}}A_{\mu}^{8}+\frac{1}{2\sqrt{6}}A_{\mu}^{15}+\frac{1}{2\sqrt{10}}A_{\mu}^{24}+YB^{0}_{\mu} \\
\\
D_{\mu}^{3} = &  -\frac{1}{\sqrt{3}}A_{\mu}^{8}+\frac{1}{2\sqrt{6}}A_{\mu}^{15}+\frac{1}{2\sqrt{10}}A_{\mu}^{24}+YB^{0}_{\mu} \\
\\
D_{\mu}^{4} = &   -\frac{3}{2\sqrt{6}}A_{\mu}^{15}+\frac{1}{2\sqrt{10}}A_{\mu}^{24}+YB^{0}_{\mu} \\
\\
D_{\mu}^{5} = & -\frac{2}{\sqrt{10}}A_{\mu}^{24}+YB^{0}_{\mu}
\end{array}
\end{equation} 
The off-diagonal entires can be put as $B_{\mu}^{\alpha\beta}=\frac{1}{\sqrt{2}}(A_{\mu}^{\alpha}\pm iA_{\mu}^{\beta})$ with $\alpha,\beta=1,2,3,4,5$, $\alpha\neq\beta$. Obviously, the off-diagonal entries correspond either to charged bosons (if $\alpha=5$ or $\beta=5$), or to neutral bosons (if simultaneously $\alpha\neq5$ and $\beta\neq5$). That means there are no exotic electric charges allowed by this model, since all the gauge bosons exhibit only $0,\pm$e charges.

The SSB is achieved by means of an appropriate scalar sector consisting of the following five scalar quintuplets

\begin{equation}
\phi^{(k)}=\left(\begin{array}{l}
\phi_{1}^{(k)}\\
\phi_{2}^{(k)}\\
\phi_{3}^{(k)}\\
\phi_{4}^{(k)}\\
\phi_{5}^{(k)}
\end{array}\right)\sim(\boldsymbol{1},\boldsymbol{5},-\frac{1}{5}),  k=1,\ldots,4\qquad \phi^{(5)}=\left(\begin{array}{l}
\phi_{1}^{(5)}\\
\phi_{2}^{(5)}\\
\phi_{3}^{(5)}\\
\phi_{4}^{(5)}\\
\phi_{5}^{(5)}
\end{array}\right)\sim(\boldsymbol{1},\boldsymbol{5},\frac{4}{5})\qquad
\end{equation}
developing each of them its own vacuum expectation value (VEV), in the manner $\left\langle\phi^{(i)}\right\rangle=\eta_{i}$V, due to a set of real parameters ($\eta_{i}\in (0,1)$) once a unique overall scale V in the model is assumed. The parameters can be grouped into a $5 \times 5$ diagonal matrix ($\eta$) whose entries obey (according to the general method \cite{5}) a restrictive  trace condition $\text{Tr}\left(\eta^{2}\right)$=1, that is $\eta^{2}_{1}+\eta^{2}_{2}+\eta^{2}_{3}+\eta^{2}_{4}+\eta^{2}_{5}=1$, so that the relation among all five VEVs $\left\langle\phi^{(1)}\right\rangle^{2}+\left\langle\phi^{(2)}
\right\rangle^{2}+\left\langle\phi^{(3)}\right\rangle^{2}+\left
\langle\phi^{(4)}\right\rangle^{2}+\left\langle\phi^{(5)}\right\rangle^{2}=V^{2}$ holds. For our purpose we have employed \cite{1} the following parameter matrix
\begin{equation}
\eta^{2}=\text{Diag}\left(\frac{1-a}{3},\frac{1-a}{3},\frac{1-a}{3},\frac{a-b}{2},\frac{a+b}{2}\right)
\end{equation}  
which fulfill the trace requirement. It also split the VEVs, as one can tune $a,b \rightarrow 0$ (very small), so that $\left\langle\phi^{(1)}\right\rangle$,$\left\langle\phi^{(2)}\right\rangle$,$\left\langle\phi^{(3)}\right\rangle\sim V$ and $\left\langle\phi^{(4)}\right\rangle$,$\left\langle\phi^{(5)}\right\rangle \sim v_{SM}$.  

The detailed procedure and the resulting Higgs spectrum are presented in Appendix B in Ref.\cite{1}.

\section{Boson mass spectrum}
\label{sec:3}

According to the general method \cite{5}, with the above parameter choice, one gets (once the SSB is achieved) the following mass matrix for the Hermitian bosons that mediate the weak interactions: 

\begin{equation}
M^{2}=\left(\begin{array}{cccc}
M^{2}(Z^{\prime\prime\prime}) & 0 & 0 & 0\\
\\
0 & M^{2}(Z^{\prime\prime}) & 0 & 0\\
\\
0 & 0 & 0 & M^{2}_{2\times2}(Z,Z^{\prime})
\end{array}\right)
\end{equation}

As expected, two of the heavier bosons are completely decoupled, only one of them ($Z^{\prime}$) mixes with the SM neutral $Z$ boson. So, only the $2 \times 2$ mass matrix
\begin{equation}
M^{2}_{2 \times 2}(Z,Z^{\prime})=\frac{m^{2}}{2}\left(\begin{array}{cc}
\frac{1}{3}\left(1+\frac{7}{2}a-\frac{9}{2}b\right) & \frac{1}{\sqrt{15}\cos\theta}\left(1-\frac{5}{2}a+\frac{3}{2}b\right)\\
\\
\frac{1}{\sqrt{15}\cos\theta}\left(1-\frac{5}{2}a+\frac{3}{2}b\right) & \frac{1}{5\cos^{2}\theta}\left(1+\frac{15}{2}a+\frac{15}{2}b\right)
\end{array}\right)
\end{equation}
goes actually through the diagonalization procedure. Here $\theta$ stands for the rotation angle of a generalized Weinberg transformation (see sec.5 in Ref.\cite{5}) that separates the massless electromagnetic direction in the parameter space. It is connected to the SM Weinberg angle ($\theta_{W}$) in our particular 3-5-1 model \cite{1} in the manner $\sin\theta=2\sqrt{\frac{2}{5}}\sin\theta_{W}$.

Now, one has to enforce precisely the mass $m^{2}a/\cos^{2}\theta_{W}$ of the SM neutral boson $Z$  ($\simeq91.2$ GeV \cite{53}) as an eigenvalue of the matrix in eq.(10). In our parametrization, $M(W^{\pm})=m\sqrt{a}$ ($\simeq80.4$ GeV \cite{53}), with the notation $m^{2}=\frac{1}{4}g^{2}V^{2}$ used throughout the proceedings.

Hence, one gets a restriction \cite{1} on the two free parameters $a$ and $b$, namely   
\begin{equation}
\left(a\tan^{2}\theta_W+b\right)^2=0
\end{equation}

Under these circumstances the boson mass matrix $M^{2}_{2\times2}(Z,Z^{\prime})$ to be diagonalized becomes the one-parameter matrix

\begin{equation}
\frac{m^{2}}{2}\left(\begin{array}{cc}
\frac{1}{3}\left[1+a\frac{(7+2\sin^{2}\theta_{W})}{2(1-\sin^{2}\theta_{W})}\right] & \frac{1}{\sqrt{3(5-8\sin^{2}\theta_{W})}} \left[1-a\frac{(5-2\sin^{2}\theta_{W})}{2(1-\sin^{2}\theta_{W})}\right]\\
\\
\frac{1}{\sqrt{3(5-8\sin^{2}\theta_{W})}} \left[1-a\frac{(5-2\sin^{2}\theta_{W})}{2(1-\sin^{2}\theta_{W})}\right] & \frac{1}{5-8\sin^{2}\theta_{W}}\left[1+a\frac{(15-30\sin^{2}\theta_{W})}{2(1-\sin^{2}\theta_{W})}\right]
\end{array}\right)
\end{equation}

The diagonalization of the matrix in eq.(9) is performed simply by employing the following $SO(4)$ matrix, 

\begin{equation}
\omega=\frac{1}{2\sqrt{2}\cos\theta_{W}}\left(\begin{array}{cccc}
1 & 0 & 0 &0 \\
0 & 1 & 0 &0 \\ 
0 & 0 & -\sqrt{3} & \sqrt{5-8\sin^{2} \theta_{W}} \\
0 & 0 & -\sqrt{5-8\sin^{2} \theta_{W}} & -\sqrt{3} \end{array} \right),
\end{equation}
in the manner
\begin{equation}
\omega M^{2} \omega^{T}=\text{Diag}\left[M^{2}(Z^{\prime\prime\prime}),M^{2}(Z^{\prime\prime}), M^{2}(Z), M^{2}(Z^{\prime})\right].
\end{equation}

The three new eigenvalues, specific to our 3-5-1 model, are now in order:
 
\begin{equation}
M^{2}(Z^{\prime})= m^{2}\left[\frac{4}{3}\left(\frac{1-\sin^{2}\theta_{W}}{5-8\sin^{2}\theta_{W}}\right)-\frac{a}{3}\frac{(5-10\sin^{2}\theta_{W}-4\sin^{4}\theta_{W})}{(5-8\sin^{2}\theta_{W})(1-\sin^{2}\theta_{W})}\right]
\end{equation}

\begin{equation}
M^{2}(Z^{\prime\prime})=M^{2}(Z^{\prime\prime\prime})=m^{2}\left(\frac{2}{3}\right)(1-a)
\end{equation}
all of them much heavier than $Z$, as it was shown in Ref.\cite{1} if one considers the parameter $a$ very small, say its order of magnitude $\emph{O}(10^{-3})$ or smaller. Such a tuning ensures an overall scale $V$ around 10 TeV or higher, according to $v_{SM}=\sqrt{a}V$. 

\section{Weak charges}
\label{sec:4}

Once we identified the $\omega$ matrix, all the weak charge operators $Q^{\rho}(Z_{\hat{i}})$ can be computed, according to the prescriptions of the general method Ref.\cite{5}, as: 
\begin{equation}
Q^{\rho}(Z_{\hat{i}})=g\left[D_{\hat{k}}^{\rho}-\nu_{\hat{k}}\left(D^{\varrho}\nu\right)(1-\cos\theta)-\nu_{\hat{k}}\frac{g^{\prime}}{g}Y^{\rho}\sin\theta\right]\omega_{\cdot\hat{i}}^{\hat{k}\cdot}
\end{equation}
where the versors $\nu_{\hat{k}}$ are associated to the Hermitian diagonal generators of the gauge group. As it was proved in Ref.\cite{1}, in order to avoid exotic electric charges, one must select for the model at hand $\nu_{n^{2}-1}=\nu_{24}=1$ and simultaneously impose the vanishing of all the other three versors $\nu_{3}=\nu_{8}=\nu_{15}=0$. In a way, one can say that these versors properly discriminate among the various models based on the same gauge group. 

Thus, in our particular 3-5-1 model, the neutral charge operators become:
\begin{equation}
Q^{\rho}(Z)=\frac{e}{s_{w}c_{w}}\left[-\sqrt{\frac{3}{8}}T_{15}^{\rho}+\sqrt{\frac{5-8s^{2}_{W}}{8}}\left(T_{24}^{\rho}\sqrt{\frac{5-8s^{2}_{W}}{5}}-\frac{2\sqrt{2}s^{2}_{W}}{\sqrt{5-8s^{2}_{W}}}Y^{\rho}\right)\right],
\end{equation}

\begin{equation}
Q^{\rho}(Z^{\prime})=\frac{e}{s_{w}c_{w}}\left[-\sqrt{\frac{5-8s^{2}_{W}}{8}}T_{15}^{\rho}-\sqrt{\frac{3}{8}}\left(T_{24}^{\rho}\sqrt{\frac{5-8s^{2}_{W}}{5}}-\frac{2\sqrt{2}s^{2}_{W}}{\sqrt{5-8s^{2}_{W}}}Y^{\rho}\right)\right],
\end{equation}

\begin{equation}
Q^{\rho}(Z^{\prime\prime})=gT^{\rho}_{8}=\frac{e}{2\sqrt{3}s_{W}}\textrm{Diag}(1,1,-2,0,0),
\end{equation}

\begin{equation}
Q^{\rho}(Z^{\prime\prime\prime})=gT^{\rho}_{3}=\frac{e}{2s_{W}}\textrm{Diag}(1,-1,0,0,0),
\end{equation}
where we made use of the notations $s_{W}=\sin\theta_{W}$ and $c_{W}=\cos\theta_{W}$ along with the identification $e=g\sin\theta_{W}$ (once we established that the coupling $g$ of the $SU(5)_{L}$ is identical with $g$ of  the $SU(2)_{L}$ in the SM). 

Now, with a little algebra, the neutral charges are inferred  straightforwardly for all the irreducible representations in our model. We opt to express these charges, as usual, in $e/2s_{W}c_{W}$ units in order to easily compare them to the well-known SM predicted values. 

The resulting couplings with the SM neutral vector boson $Z$ are

\begin{equation}
Q^{(\boldsymbol{5},-\frac{1}{5})}\left(Z\right)=\left(\begin{array}{ccccc}
0\\
 & 0\\
 &  & 0\\
 &  &  & 1\\
 &  &  &  & -1+2s^{2}_{W}
\end{array}\right)
\end{equation}
for the lepton sector, and

\begin{equation}
Q^{(\boldsymbol{5},\frac{7}{15})}\left(Z\right)=\left(\begin{array}{ccccc}
-\frac{4s^{2}_{W}}{3}\\
 & -\frac{4s^{2}_{W}}{3}\\
 &  & -\frac{4s^{2}_{W}}{3}\\
 &  &  & 1-\frac{4s^{2}_{W}}{3}\\
 &  &  &  & -1+\frac{2s^{2}_{W}}{3}
\end{array}\right),
\end{equation}

\begin{equation}
Q^{(\boldsymbol{5^{*}},-\frac{2}{15})}\left(Z\right)=\left(\begin{array}{ccccc}
\frac{2s^{2}_{W}}{3}\\
 & \frac{2s^{2}_{W}}{3}\\
 &  & \frac{2s^{2}_{W}}{3}\\
 &  &  & -1+\frac{2s^{2}_{W}}{3}\\
 &  &  &  & 1-\frac{4s^{2}_{W}}{3}
\end{array}\right)
\end{equation}
for the quark sector, respectively.

For $Z^{\prime}$ neutral vector boson, the couplings are - up to a factor $\frac{\sqrt{3}}{\sqrt{5-8s^{2}_{W}}}$ - yield

\begin{equation}
Q^{(\boldsymbol{5},-\frac{1}{5})}\left(Z^{\prime}\right)=\left(\begin{array}{ccccc}
-\frac{2c^{2}_{W}}{3}\\
 & -\frac{2c^{2}_{W}}{3}\\
 &  & -\frac{2c^{2}_{W}}{3}\\
 &  &  & 1-2s^{2}_{W}\\
 &  &  &  & 1-2s^{2}_{W}
\end{array}\right)
\end{equation}
for the lepton sector, and

\begin{equation}
Q^{(\boldsymbol{5},\frac{7}{15})}\left(Z^{\prime}\right)=\left(\begin{array}{ccccc}
-\frac{2(1-3s^{2}_{W})}{3}\\
 & -\frac{2(1-3s^{2}_{W})}{3}\\
 &  & -\frac{2(1-3s^{2}_{W})}{3}\\
 &  &  & 1-\frac{2s^{2}_{W}}{3}\\
 &  &  &  & 1-\frac{2s^{2}_{W}}{3}
\end{array}\right),
\end{equation}

\begin{equation}
Q^{(\boldsymbol{5^{*}},-\frac{2}{15})}\left(Z^{\prime}\right)=\left(\begin{array}{ccccc}
\frac{2(1-2s^{2}_{W})}{3}\\
 & \frac{2(1-2s^{2}_{W})}{3}\\
 &  & \frac{2(1-2s^{2}_{W})}{3}\\
 &  &  & -1+\frac{4s^{2}_{W}}{3}\\
 &  &  &  & -1+\frac{4s^{2}_{W}}{3}
\end{array}\right)
\end{equation}
for the quark sector, respectively.

The results for the neutral charges of the SM fermions are summarized in Table 1. It is now something of an evidence that the couplings connecting any SM-fermion to the neutral SM-vector boson ($Z$) are utterly recovered, meaning that the SM is not altered at all at tree level. At the same time, the heavier $Z^{\prime\prime}$ and $Z^{\prime\prime\prime}$, being completely decoupled, exhibit no interactions with the SM fermions. That means the two bosons do not interfere with the established SM phenomenology. Only $Z^{\prime}$ could eventually somehow influence the SM phenomenology. Therefore, it deserves a distinct work in which the necessary corrections are properly performed in order to provide us with some restrictions regarding the parameters of this model.  

\begin{table}[t]
  \begin{center}
    \caption{Couplings of the SM fermions}
    \label{tab:table1}
    \begin{tabular}{|l|c|c|c|c|} 
    \hline      
   couplings $\left(\times\frac{e}{2s_{W}c_{W}}\right)$ & \textbf{$Z$} & \textbf{$Z^{\prime}$} & \textbf{$Z^{\prime \prime}$} & \textbf{$Z^{\prime \prime \prime}$}\\
     \hline
     \hline
     
      $e_{L}$, $\mu_{L}$, $\tau_{L}$ & $-1+2s^{2}_{W}$ & $\frac{(1-2s^{2}_{W})\sqrt{3}}{\sqrt{5-8s^{2}_{W}}}$ & 0 & 0 \\
      $\nu_{eL}$, $\nu_{\mu L}$, $\nu_{\tau L}$  & 1 & $\frac{(1-2s^{2}_{W})\sqrt{3}}{\sqrt{5-8s^{2}_{W}}}$ & 0 & 0\\
      $e_{R}$, $\mu_{R}$, $\tau_{R}$ & $2s^{2}_{W}$ & $\frac{-2\sqrt{3}s^{2}_{W}}{\sqrt{5-8s^{2}_{W}}}$ & 0 & 0 \\
      $\nu_{eR}$, $\nu_{\mu R}$, $\nu_{\tau R}$  & 0 & 0 & 0 & 0\\
      \hline
      $u_{L}$, $c_{L}$ & $1-\frac{4}{3}s^{2}_{W}$ & $\frac{(-1+\frac{4}{3} s^{2}_{W})\sqrt{3}}{\sqrt{5-8s^{2}_{W}}}$ & 0 & 0 \\
      $t_{L}$ & $1-\frac{4}{3}s^{2}_{W}$ & $\frac{(1-\frac{2}{3} s^{2}_{W})\sqrt{3}}{\sqrt{5-8s^{2}_{W}}}$ & 0 & 0 \\
      $d_{L}$, $s_{L}$  & $-1+\frac{2}{3}s^{2}_{W}$ & $\frac{(-1+\frac{4}{3} s^{2}_{W})\sqrt{3}}{\sqrt{5-8s^{2}_{W}}}$ & 0 & 0\\
      $b_{L}$  & $-1+\frac{2}{3}s^{2}_{W}$ & $\frac{(1-\frac{2}{3} s^{2}_{W})\sqrt{3}}{\sqrt{5-8s^{2}_{W}}}$ & 0 & 0\\
      $u_{R}$, $c_{R}$, $t_{R}$ & $-\frac{4}{3}s^{2}_{W}$ & $\frac{4s^{2}_{W}}{\sqrt{3}\sqrt{5-8s^{2}_{W}}}$ & 0 & 0 \\
      $d_{R}$, $s_{R}$, $b_{R}$  & $\frac{2}{3}s^{2}_{W}$ & $-\frac{2s^{2}_{W}}{\sqrt{3}\sqrt{5-8s^{2}_{W}}}$ & 0 & 0\\
      \hline
      \hline
    \end{tabular}
  \end{center}
\end{table}

For the the heavier $Z^{\prime\prime}$ and $Z^{\prime\prime\prime}$ neutral vector bosons the computation is much simpler, since their couplings are connected only to their associated diagonal generators. That is, only $T_{3}$ accounts for $Z^{\prime\prime\prime}$ couplings and $T_{8}$ accounts for $Z^{\prime\prime}$ couplings, with no admixture at all. The resulting values are summarized in Table 2, once the explicit expressions are computed in the following.

In the case of the $Z^{\prime\prime}$, the couplings are inferred as:

\begin{equation}
Q^{(\boldsymbol{5},-\frac{1}{5})}\left( Z^{\prime\prime}\right)=\frac{c_{W}}{\sqrt{3}}\left(\begin{array}{ccccc}
1\\
 & 1\\
 &  & -2\\
 &  &  & 0\\
 &  &  &  & 0
\end{array}\right)
\end{equation}
for the lepton sector, and

\begin{equation}
Q^{(\boldsymbol{5},\frac{7}{15})}\left( Z^{\prime\prime}\right)=\frac{c_{W}}{\sqrt{3}}\left(\begin{array}{ccccc}
1\\
 & 1\\
 &  & -2\\
 &  &  & 0\\
 &  &  &  & 0
\end{array}\right)
\end{equation}

\begin{equation}
Q^{(\boldsymbol{5^{*}},-\frac{2}{15})}\left( Z^{\prime\prime}\right)=\frac{c_{W}}{\sqrt{3}}\left(\begin{array}{ccccc}
-1\\
 & -1\\
 &  & 2\\
 &  &  & 0\\
 &  &  &  & 0
\end{array}\right)
\end{equation}
for the quark sector, respectively.

The couplings for $Z^{\prime\prime\prime}$ yield:

\begin{equation}
Q^{(\boldsymbol{5},-\frac{1}{5})}\left( Z^{\prime\prime\prime}\right)=c_{W}\left(\begin{array}{ccccc}
1\\
 & -1\\
 &  & 0\\
 &  &  & 0\\
 &  &  &  & 0
\end{array}\right)
\end{equation}
for the lepton sector, and

\begin{equation}
Q^{(\boldsymbol{5},\frac{7}{15})}\left( Z^{\prime\prime\prime}\right)=c_{W}\left(\begin{array}{ccccc}
1\\
 & -1\\
 &  & 0\\
 &  &  & 0\\
 &  &  &  & 0
\end{array}\right)
\end{equation}

\begin{equation}
Q^{(\boldsymbol{5^{*}},-\frac{2}{15})}\left( Z^{\prime\prime\prime}\right)=c_{W}\left(\begin{array}{ccccc}
-1\\
 & 1\\
 &  & 0\\
 &  &  & 0\\
 &  &  &  & 0
\end{array}\right)
\end{equation}
for the quark sector, respectively. 
 
\begin{table}[t]
  \begin{center}
    \caption{Couplings of the non-SM fermions}
    \label{tab:table2}
    \begin{tabular}{|l|c|c|c|c|}
    \hline       
   couplings $\left(\times\frac{e}{2s_{W}c_{W}}\right)$ & \textbf{$Z$} & \textbf{$Z^{\prime}$} & \textbf{$Z^{\prime \prime}$} & \textbf{$Z^{\prime \prime \prime}$}\\
     \hline
     \hline
     
      $N_{eL}$, $N_{\mu L}$, $N_{\tau L}$ & 0 & $\frac{2c^{2}_{W}}{\sqrt{3}\sqrt{5-8s^{2}_{W}}}$ & $-\frac{2c_{W}}{\sqrt{3}}$ & 0 \\
      $N^{\prime}_{eL}$, $N^{\prime}_{\mu L}$, $N^{\prime}_{\tau L}$  & 0 & $\frac{2c^{2}_{W}}{\sqrt{3}\sqrt{5-8s^{2}_{W}}}$ & $\frac{c_{W}}{\sqrt{3}}$ & $-c_{W}$\\
      $N^{\prime\prime}_{eL}$, $N^{\prime\prime}_{\mu L}$,   $N^{\prime\prime}_{\tau L}$  & 0 & $\frac{2c^{2}_{W}}{\sqrt{3}\sqrt{5-8s^{2}_{W}}}$ & $\frac{c_{W}}{\sqrt{3}}$ & $c_{W}$\\
      $N_{eR}$, $N_{\mu R}$, $N_{\tau R}$ & 0 & 0 & 0 & 0 \\
      $N^{\prime}_{eR}$, $N^{\prime}_{\mu R}$, $N^{\prime}_{\tau R}$  & 0 & 0 & 0 & 0\\
      $N^{\prime\prime}_{eR}$, $N^{\prime\prime}_{\mu R}$, $N^{\prime\prime}_{\tau R}$  & 0 & 0 & 0 & 0\\
      \hline         
      $U_{1L}$  & $-\frac{4}{3}s^{2}_{W}$ & $-\frac{2(1-3s^{2}_{W})}{\sqrt{3}\sqrt{5-8s^{2}_{W}}}$ & $-\frac{2c_{W}}{\sqrt{3}}$ & 0 \\
      
      $D_{2L}$, $D_{3L}$ & $\frac{2}{3}s^{2}_{W}$ & $\frac{2(1-2s^{2}_{W})}{\sqrt{3}\sqrt{5-8s^{2}_{W}}}$ & $\frac{2c_{W}}{\sqrt{3}}$ & 0 \\
     
      $U^{\prime}_{1L}$  & $-\frac{4}{3}s^{2}_{W}$ & $-\frac{2(1-3s^{2}_{W})}{\sqrt{3}\sqrt{5-8s^{2}_{W}}}$ & $\frac{c_{W}}{\sqrt{3}}$ & $-c_{W}$ \\
     
      $D^{\prime}_{2L}$, $D^{\prime}_{3L}$  & $\frac{2}{3}s^{2}_{W}$ & $\frac{2(1-2s^{2}_{W})}{\sqrt{3}\sqrt{5-8s^{2}_{W}}}$  & $-\frac{c_{W}}{\sqrt{3}}$ & $c_{W}$ \\
      
      $U^{\prime\prime}_{1L}$ & $-\frac{4}{3}s^{2}_{W}$ & $-\frac{2(1-3s^{2}_{W})}{\sqrt{3}\sqrt{5-8s^{2}_{W}}}$ &  $\frac{c_{W}}{\sqrt{3}}$ & $c_{W}$ \\

      $D^{\prime\prime}_{2L}$, $D^{\prime\prime}_{3L}$ & $\frac{2}{3}s^{2}_{W}$ & $\frac{2(1-2s^{2}_{W})}{\sqrt{3}\sqrt{5-8s^{2}_{W}}}$  & $-\frac{c_{W}}{\sqrt{3}}$ & $-c_{W}$ \\
     
      $U_{1R}$, $U^{\prime}_{1R}$, $U^{\prime\prime}_{1R}$ & $-\frac{4}{3}s^{2}_{W}$ & $\frac{4s^{2}_{W}}{\sqrt{3}\sqrt{5-8s^{2}_{W}}}$ & 0 & 0 \\
      $D_{2R}$, $D_{3R}$, $D^{\prime}_{2R}$, $D^{\prime}_{3R}$,  $D^{\prime\prime}_{2R}$, $D^{\prime\prime}_{3R}$ & $\frac{2}{3}s^{2}_{W}$ & $-\frac{2s^{2}_{W}}{\sqrt{3}\sqrt{5-8s^{2}_{W}}}$ & 0 & 0 \\
        \hline
        \hline
    \end{tabular}
  \end{center}
\end{table}

Let's turn now to the fermion singlets. Their weak charges are computed in a simpler way, since one takes into account only $Y$ and none of the $T$ generators. Hence, there are no interactions at all for the right-handed singlets with the heavier $Z^{\prime\prime}$ and $Z^{\prime\prime\prime}$. At the same time, as expected, all the neutral right-handed fermions have no weak interactions regardless the SM or non-SM bosons.  

For the sake of completeness we display below the weak interactions of the right handed representations explicitly. In the case of charged leptons we get the following weak couplings
\begin{equation}
Q^{(\boldsymbol{1},-1)}\left( Z\right)=2s^{2}_{W}, \quad
Q^{(\boldsymbol{1},-1)}\left( Z^{\prime}\right)=-\frac{2\sqrt{3}s^{2}_{W}}{\sqrt{5-8s^{2}_{W}}},
\end{equation}
while all kinds of right-handed neutrinos are sterile  
\begin{equation}
Q^{(\boldsymbol{1},0)}\left( Z\right)=0, \quad
Q^{(\boldsymbol{1},0)}\left( Z^{\prime}\right)=0,
\end{equation}
as expected.

In the quark sector, the right-handed up-type quarks couplings yield
\begin{equation}
Q^{(\boldsymbol{1},2/3)}\left( Z\right)=-\frac{4s^{2}_{W}}{3},\quad
Q^{(\boldsymbol{1},2/3)}\left( Z^{\prime}\right)=\frac{4s^{2}_{W}}{\sqrt{3(5-8s^{2}_{W})}},
\end{equation}
while the right-handed down-type quarks interacts weakly in the manner
\begin{equation}
Q^{(\boldsymbol{1},-1/3)}\left( Z\right)=\frac{2s^{2}_{W}}{3},\quad
Q^{(\boldsymbol{1},-1/3)}\left( Z^{\prime}\right)=-\frac{2s^{2}_{W}}{\sqrt{3(5-8s^{2}_{W})}}.
\end{equation}

\section{Concluding remarks} 
\label{sec:5}
As one can easily observe from the results derived and presented above, the SM-fermions preserve their couplings predicted by the SM. Moreover, they have no couplings at all with the new bosons $Z^{\prime\prime}$ and $Z^{\prime\prime\prime}$, but only with $Z^{\prime}$ ($\geq5.1$ TeV \cite{50}) whose influence (due to the mixing with $Z$ in the diagonalization procedure) can (in a future work) be estimated. If worked out properly such corrections could enforce restrictions on the scale of the model and other parameters as well (for instance, the appropriate Yukawa couplings in the neutrino sector or quark sector). At the same time, a particular feature of our 3-5-1 model is that SM-boson $Z$ has vector interactions with all heavier fermions (other than SM-fermions), while $Z^{\prime}$ makes no distinction on the electric charge basis when it comes to the left-handed fermions in the same doublet from the SM. At the same time, this kind of models supplies a plethora of suitable candidates for the cold Dark Matter, since a lot of neutral femions, neutral scalars and even neutral vector bosons are not interacting with ordinary matter particles - as the new scale $V$ is, properly speaking, completely decoupled from the SM's scale. Thus, such eluding particles could naturally have escaped to any observation up to date and their relic density can be in principle estimated.  

Having said all these, we consider that the particular 3-5-1 electro-weak unification with no exotic charges is a viable SM-extension candidate, so that it truly deserves attention and further study for deepening its phenomenological investigation.

\end{document}